# An Approach to 2D Photodetectors - Absorption of Bloch Surface Waves Using Nanometric Thin Graphene Layers

R. Dubey[1]*, B. V. Lahijani[1], M. Marchena[2], V. Pruneri[2] and H. P. Herzig[1]

[1]Optics & Photonics Technology Laboratory (OPT), École Polytechnique Fédérale de Lausanne (EPFL), rue de la Maladière 71b, CH-2002 Neuchâtel, Switzerland

[2]ICFO–Institut de Ciències Fotòniques, The Barcelona Institute of Science and Technology, 08860 Castelldefels (Barcelona), Spain

*Corresponding author: richa.dubey@epfl.ch

**Abstract:** A dielectric multilayer platform has been investigated as a foundation for two-dimensional optics. In this paper, we present, to the best of our knowledge, the first experimental demonstration of absorption of Bloch surface waves in the presence of graphene layers. Graphene layers have been grown previously *via* CVD on Cu foils and transferred layer by layer by PMMA-wet transfer method. We exploit total internal reflection configuration and multi-heterodyne scanning near-field optical microscopy as a far-field coupling method and near-field characterization tool, respectively. The absorption is quantified in terms of propagation lengths of Bloch surface waves. A significant drop (about 17 times shorter) in the propagation length of surface waves is observed in the presence of single layer of graphene.

## 1. Introduction

A dielectric multilayer platform, sustaining electromagnetic surface waves, has been studied extensively for sensing application [1, 2]. The electromagnetic surface waves propagating at the interface between dielectric multilayer stack and external medium are called Bloch Surface Waves (BSWs) [3, 4]. The modes propagating on the multilayer interface decays exponentially inside the multilayers due to presence of photonics band gap. At the same time modes are not allowed to propagate in the external medium, which is air in this study, because of total internal reflection. This salient propagation mechanism keeps the modes bounded close to the multilayer surface. The optical field confinement in the vertical direction (perpendicular to the multilayer interface) opens the ways to study this platform for as a foundation for two-dimensional (2D) optics. The absorption in the material, surface scattering, and leakage into the multilayer because of prism coupling [5, 6] are the main sources of the decay of the surface mode amplitude along the propagation direction. However, in the present system, material absorption and surface scattering do not have a significant contribution to the losses. This is because of low absorption of dielectric materials and the low roughness of the multilayer surface [7]. Therefore, the BSWs are mainly decaying into the multilayers due to light leakage in the prism coupling system [8, 9, 10]. The decay coefficient decreases exponentially with the number of periods of multilayer stack [3]. The present design of multilayer is optimized to achieve longer propagation lengths [8]. The BSW platform provides the possibility of ample choice of constituent material, in comparison to surface plasmon polaritons (SPPs), unless material is transparent at the operating wavelengths. In addition, the maximum field amplitude can be tuned at the surface by tailoring the thickness of the topmost layer. This ability to tune local field confinement is attractive for sensing applications [11, 12]. One of the prominent advantages of platform concept is standard wafer-scale production for the fabrication of thin film multilayers.

Aforementioned advantages and features make the BSW platform an interesting candidate for the development of 2D optical systems. To this aim, several 2D optical components based on BSWs have been studied theoretically and experimentally. They include ridge waveguides, lenses, disk resonators, BSWs reflectors, Bessel-like beams, subwavelength focusers, phase-shifted Bragg gratings, and grating couplers [13, 6, 14-30]. Further, the tunable planar optical components on BSW platform have been demonstrated [31, 32]. However, the absorption of propagating BSWs on the top of multilayer platform has not been studied yet.

In this paper, the absorption of BSWs, using graphene layers, has been demonstrated experimentally. We use graphene layer as a device layer on the top of multilayer platform. The electro optic properties of graphene can be exploited to design active 2D optical components on BSW platform. The existence of BSWs on graphene based multilayer platform has been proven [33]. We study the propagation properties of single and bilayer of graphene. The thicknesses of single and bilayer of graphene are about 0.3 nm and 0.6 nm, respectively. The deposition of graphene layers on BSW platform is presented in 'Materials and Methods' section.

Graphene is a two-dimensional single layer of carbon atoms forming a hexagonal lattice. Graphene exhibits exceptional electrical and optical properties. Because of being a zero-band gap semiconductor, the electronic properties of graphene have drawn much attention in research. The high-speed electronics devices have been already proposed, thanks to ultrafast response of the graphene [34]. Apart from the electronics properties of Graphene, optical properties show equal interest because of strong interaction of graphene with light for broad band of wavelengths. The optical conductivity of graphene (from visible to infrared wavelengths) leads to absorption of 2.3% of light in case of monolayer [33]. These optical properties made graphene attractive for photonics and optoelectronics devices such as photovoltaic devices, optical modulator and photodetectors, etc [35]. One other great advantage of graphene is its electro optic effect where the optical properties of graphene can be changes either by electron doping or nonlinear effect [36]. With the advantage of tunable optical absorption and ultrafast response, Graphene has proven a promising candidate for on-chip integrated silicon photonics. For example, Graphene-based photodetector integrated on Si waveguide have been studied by S. Schuler et al [35].

## 2. Materials and Methods

The dielectric multilayer platform presented in this study made up of six periods of alternating high index and low index layers. The low index and high index material layers are silicon dioxide ($SiO_2$) and silicon nitride ($SiN_x$) with respective a refractive index of 1.45 and 1.79 around wavelength, $\lambda$ = 1550 nm. The thicknesses of the $SiN_x$ and $SiO_2$ layers are 283 nm and 472 nm, respectively. To terminate the precocity of multilayers, 50 nm thick top layer of $SiN_x$ is deposited on the top. The top layer is addressed as a defect layer. The complete stack (periodic multilayers + defect layer) is called bare multilayer (BML) platform. The BML structure is deposited on glass wafer as shown in Fig. 1. In the present study multilayers are designed to support transverse electric (TE) polarization of incident light at telecommunication wavelengths. The multilayer platform is fabricated using Plasma-Enhanced Chemical Vapor Deposition (PECVD, PlasmaLab 80+ by Oxford Instruments) technique.

For graphene deposition, we have initially grown graphene on Cu foils of 25 μm thickness (Sigma Aldrich) using CVD (Black Magic 4-inch, AIXTRON) under the following conditions: CH$_4$:H$_2$ (1:4), 25 mbar and 10 minutes. Prior to graphene growth, Cu foil was first cleaned by rinsing in organic solvents and DI water (acetone: isopropyl alcohol: H$_2$O, 2 minutes each), and finally in 0.1M aqueous acetic acid (CH$_3$COOH) for 2 minutes to remove oxides from the Cu surface. Then, Cu foil is placed inside the CVD chamber and heated at 50°C min⁻¹ from room temperature to 1000°C under an Ar/H$_2$ flow.

After growth, graphene on Cu foil is spin-coated with PMMA. After drying it in air for 30 minutes, the whole structure is located floating on a Cu etchant solution (0.05 g/ml, ammonium persulfate) for 4 hours. To avoid etchant residues, PMMA/graphene is rinsed three times in deionized water and located on the dielectric multilayer platform. After drying it with N$_2$ and overnight in vacuum, PMMA is removed by samples immersion in acetone and isopropyl alcohol (15 minutes at each solvent). For building bilayer graphene structures on top of the dielectric multilayer, the PMMA transfer has been repeated twice [37].

To couple incident light into surface waves, we use a total internal reflection (TIR) configuration. The TIR configuration is made up of BK7-glass prism (n$_{BK7}$ = 1.501), schematic of which is presented in Fig. 1. The glass prism is required to fulfill the phase matching condition to excite BSWs. The details can be found elsewhere [10]. We illuminate the sample with a focused Gaussian beam to couple light to the BSW. At an incident angle '$\theta$' which is higher than the critical angle, the incident light is coupled to BSW at a specific wavelength. The phenomenon of BSW coupling can be seen as a reflection dip of the angular reflectance plot in the far-field (FF). We use CCD IR camera to collect the reflected light in the far-field on the other side of prism, see Fig. 1.

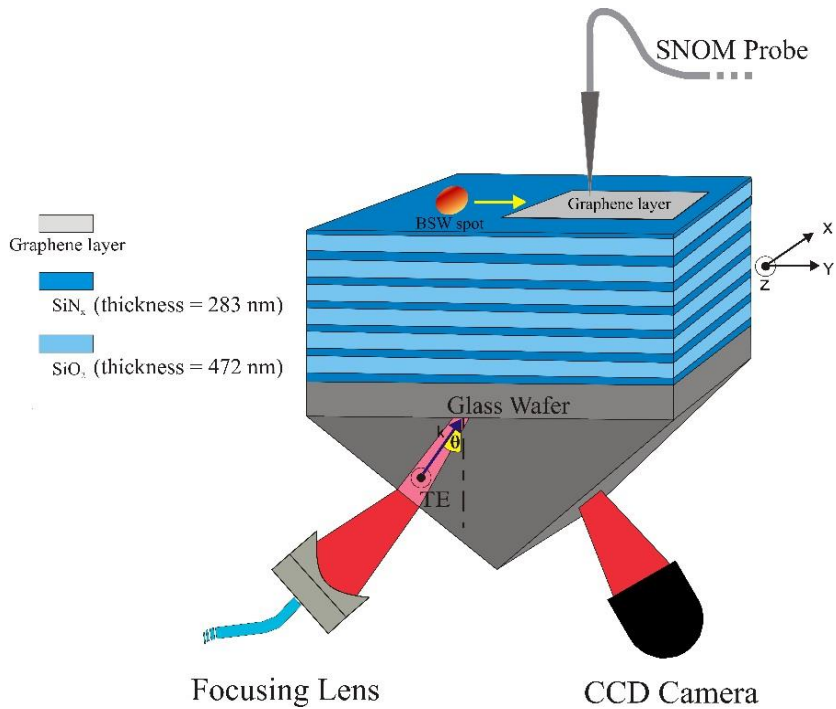

**Figure 1.** Illustration of setup of the TIR configuration for BSW coupling with dielectric multilayers deposited on a glass wafer. The SNOM probe collects the evanescent field on top of the 2D structures. On the other side of prism, there is a CCD camera to collect the light reflected in the far-field.

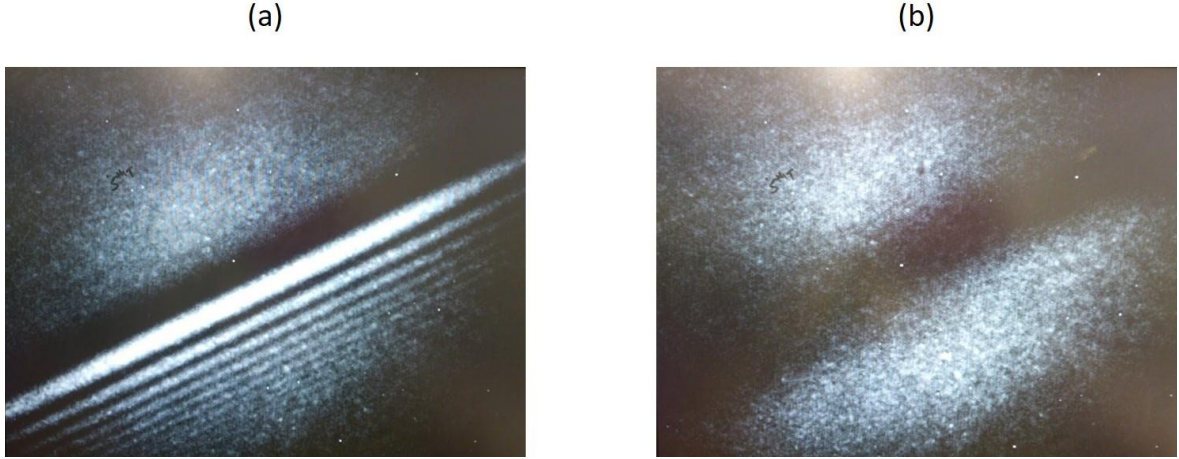

**Figure 2.** M-line pattern observed in far-field image of the reflected light captured in CCD camera in TIR coupling configuration for (**a**) Bare multilayer; (**b**) Graphene layer

For BML, the BSWs can be coupled at an incident angle $\theta$ =50.15 deg., at the corresponding wavelength $\lambda$ = 1558 nm. The specific fringe pattern can be observed in far-field, see Fig. 2(a). The fringes are produced because of interaction of m-lines and the focused spot produced by total internally reflected incident light in a prism coupling arrangement [38].

In case of BML+ graphene layer, the optimum condition of BSWs coupling is found at the same angle of incidence as BML. The nanometric thicknesses (< 1 nm) of the graphene layer could be the reason behind the same coupling angle. However, for graphene layer only broad dark band can be observed, as shown in Fig. 2(b). The dark band represents the significant absorption of propagating BSWs in the presence of graphene layer. The fact is further demonstrated by near-field experiments. The details are provided in 'Results and Discussion' section.

To perform the near-field experiments, we work with multi-heterodyne scanning near-field optical microscope (MH-SNOM) in collection mode which collects the evanescent electric field with a subwavelength aperture fiber probe. Because BSWs propagates at the interface of multilayer, near-field microscopy is an optimum tool to perform the spatial field distribution mapping locally. Further details about MH-SNOM can be found in [10].

## 3. Results and Discussion

The light is coupled evanescently into graphene layer using TIR configuration. Further, with the aid of MH-SNOM, propagation lengths of BSWs ($L_{BSW}$) for BML, graphene single layer and graphene bilayer are measured in near-field, as shown in Fig. 3. The propagation lengths are deduced by exponentially fitting the decrease of the field amplitude of the surface wave along the propagation direction [10]. The measured $L_{BSW}$ for BML, single and bilayer of graphene are approximately 1300 μm [8], 75 μm and 55 μm, respectively. From the measurement results we observe that propagation length in graphene layers (75/55 μm) are reduced drastically, roughly 17/23 times lower than bare multilayer (1.35 millimeters). This is due to strong absorption due to the presence of graphene layers.

For the comparison of absorption of propagating field, we show here the propagation of BSW without graphene layer/bare multilayer (Fig. 3(a)), with single layer of graphene (Fig. 3(b)) and with double layer of graphene (Fig. 3(c)).

We define propagation length as a distance where amplitude of the propagating field drops down to the 1/e value. It can be inferred that about 63 % of the incident light is absorbed in the span of its propagation length. We observe here that, in case of single layer of graphene 63 % of the light is absorbed in 75 μm distance in propagation direction, however bilayer of graphene depicts the absorption lengths of 55 μm.

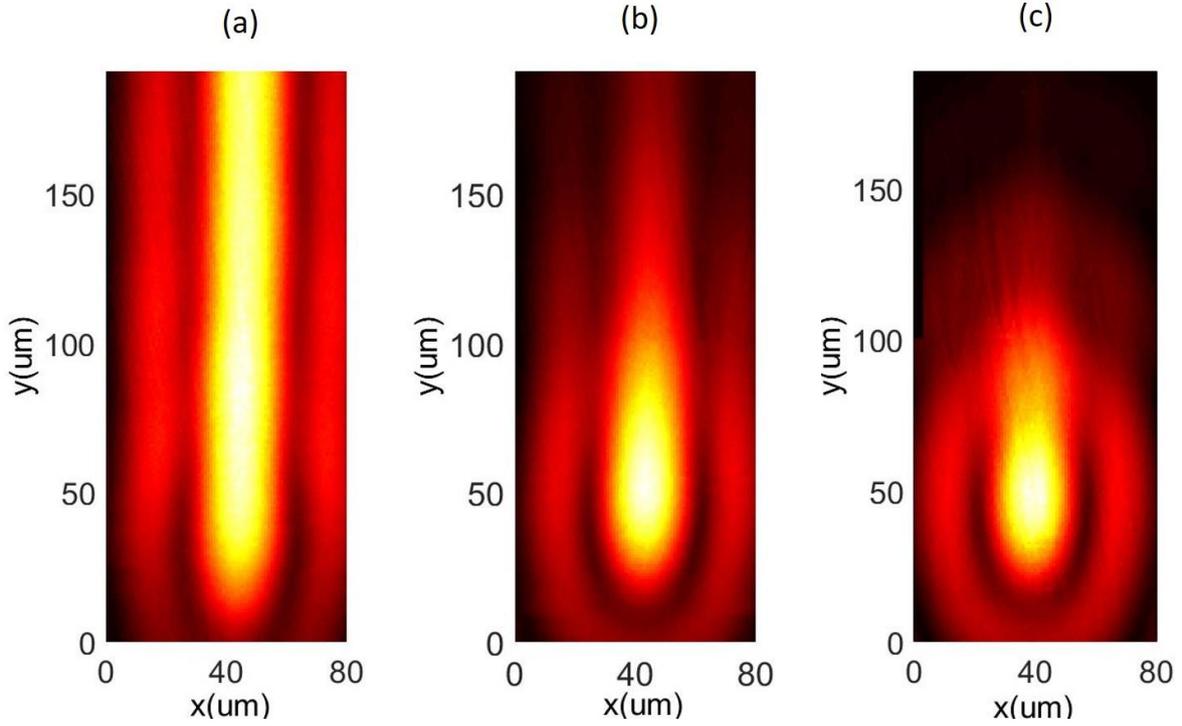

**Figure 3.** Near-field amplitude distribution acquired by MH-SNOM demonstrating BSW propagation for (**a**) Bare multilayer; (**b**) Single layer of graphene; (**c**) Bilayer of graphene

The grating coupler assisted GaInAsSb photodiode integrated on silicon waveguide has been studied somewhere else [39]. The reported responsivity of 50 μm long photodiode is 0.4A/W. From the reported results, it can be inferred that 22.5% of the incident light is absorbed (assuming 100% carrier collection efficiency). The poor coupling efficiency of the grating coupler is stated as the main reason behind the low responsivity.

However, in the present study where graphene layer is deposited on the top of BSW platform shows better absorption characteristic than GaInAsSb based integrated photodiodes. The nanometric thickness of graphene layer provides the optimum coupling of surface waves at the same coupling conditions as BML. In addition, the deposition of graphene layer on multilayer platform using standard Chemical Vapor Deposition technique provides ease of fabrication in comparison to typical Si photonics integration processes.

**4. Conclusions**

To conclude, for the first time, to the best of our knowledge, the absorption of BSWs, using graphene layers, has been demonstrated experimentally in near-field. The absorption of surface waves in the presence of graphene layers is evident from the MH-SNOM near-field measurement results. The propagation lengths of BSWs for BML and BML+ graphene layers have been compared. The results show that propagation length in presence of single graphene layer is decreased significantly (about 17 times shorter) than the bare multilayer. This is due to strong absorption in the presence of graphene layer. Our results present that the absorption does not increase proportionally when thickness of the graphene layer folds twice. Further investigations are required to understand the absorption properties of graphene, when light is when light is evanescently coupled to it, in depth. The presented absorption characteristics of graphene, make it a good candidate for BSWs based 2D photodetectors. This study provides an advancement in the direction of the development of BSWs based 2D integrated optical components.

**Acknowledgments:** We acknowledge funding from Swiss National Science Foundation (SNSF) (200020_135455).

**References**

1. Descrovi, E.; Frascella, F.; Sciacca, B.; Geobaldo, F.; Dominici, L.; Michelotti, F. Coupling of surface waves in highly defined one-dimensional porous silicon photonic crystals for gas sensing applications. *Appl. Phys. Lett.* **2007**, *91*, 241109. [10.1063/1.2824387]
2. Giorgis, F.; Descrovi, E.; Summonte, C.; Dominici, L.; Michelotti, F. Experimental determination of the sensitivity of Bloch surface waves based sensors. *Opt. Express.* **2010**, *18*, 8087–8093. [10.1364/OE.18.008087]
3. Yeh, P.; Yariv, A.; Cho, A.Y. Optical surface waves in periodic layered media. *Appl. Phys. Lett.* **1978**, *32*, 104–105. [10.1063/1.89953]
4. Liscidini , M.; Sipe, J. E. Enhancement of diffraction for biosensing applications via Bloch surface waves. *Appl. Phys. Lett.* **2007**, *91*, 253125. [10.1063/1.2826545]
5. Soboleva, I. V.; Descrovi, E.; Summonte, C.; Fedyanin, A. A.; Giorgis, F. Fluorescence emission enhanced by surface electromagnetic waves on one-dimensional photonic crystals. *Appl. Phys. Lett.* **2009**, *94*, 231122. [10.1063/1.3148671]
6. Yu, L.; Barakat, E.; Sfez, T.; Hvozdara, L.; Di Francesco, J.; Herzig, H.P. Manipulating Bloch surface waves in 2D: a platform concept-based flat lens. *Light Sci. Appl.* **2014**, *3*, 124. [10.1038/lsa.2014.5]
7. Bontempi, E.; Depero, L. E.; Sangaletti, L.; Giorgis, F.; Pirri, C. F. Growth process analysis of a-Si1-xNx:H films probed by X-ray reflectivity. *Mater. Chem. Phys.* **2000**, *66*, 172–176. [10.1016/S0254-0584(00)00338-2]
8. Dubey, R.; Barakat, E.; Häyrinen, M.; Roussey, M.; Honkanen, S.; Kuittinen, M.; Herzig, H. P. Experimental investigation of the propagation properties of Bloch surface waves on dielectric multilayer platform. *J. Eur. Opt. Soc.* **2017**, *13*, 1–9. [10.1186/s41476-016-0029-1]
9. Descrovi, E.; Sfez, T.; Dominici, L.; Nakagawa, W.; Michelotti, F.; Giorgis, F.; Herzig, H. P. Near-field imaging of Bloch surface waves on silicon nitride one-dimensional photonic crystals. *Opt. Express.* **2008**, *16*, 5453–5464. [10.1364/OE.16.005453]
10. Dubey, R. Near-field characterization of Bloch surface waves based 2D optical components. PhD Dissertation, École polytechnique fédérale de Lausanne, 2017. [infoscience.epfl.ch/record/230136]
11. Konopsky, V. N.; Alieva, E. V. Photonic crystal surface waves for optical biosensors. *Anal. Chem.* **2007**, *79*, 4729–4735. [10.1021/ac070275y]
12. Robertson, W.M.: Experimental measurement of the effect of termination on surface electromagnetic waves in one-dimensional photonic bandgap arrays. *J Light Technol.* **1999**, *17*, 2013–2017. [10.1109/50.802988]
13. Sfez, T.; Descrovi, E.; Yu, L.; Quaglio, M.; Dominici, L.; Nakagawa, W.; Michelotti, F.; Giorgis, F.; Herzig, H.P. Two-dimensional optics on silicon nitride multilayer: Refraction of Bloch surface waves. *Appl. Phys. Lett.* **2010**, *96*, 151101. [10.1063/1.3385729]

14. Dubey, R.; Lahijani, B. V.; Barakat, E.; Häyrinen, M.; Roussey, M.; Kuittinen, M.; Herzig, H. P. Near-field characterization of a Bloch-surfacewave-based 2D disk Resonator. *Opt. Lett.* **2016**, *41*, 4867-4870. [10.1364/OL.41.004867]
15. Lahijani, B. V.; Ghavifekr, H. B.; Dubey, R.; Kim, M.-S.; Vartiainen, I.; Roussey, M.; Herzig, H. P. Experimental demonstration of critical coupling of whispering gallery mode cavities on Bloch surface wave platform. *Opt. Lett.* **2017**, *42*, 5137- 5140. [10.1364/OL.42.005137]
16. Dubey, R.; Lahijani, B. V.; Häyrinen, M.; Roussey, M.; Kuittinen, M.; Herzig, H. P. Ultra-thin Bloch surface waves based reflector at telecommunication wavelength. *Photonics Res.* **2017**, *5*, 494-499. [10.1364/PRJ.5.000494]
17. Kim, M.-S.; Dubey, R.; Barakat, E.; Herzig, H. P. Nano-thin 2D axicon generating nondiffracting surface waves. Proceeding of Optical MEMS and Nanophotonics (OMN), Singapore, 2016. [10.1109/OMN.2016.7565823]
18. Kim, M.-S.; Barakat, E.; Dubey, R.; Scharf, T.; Herzig, H. P. Nondiffracting Bloch surface wave: 2D quasi-Bessel-Gauss beam. Proceeding of OSA CLEO Pacific Rim Conference, Busan, Korea, 2015. [10.1109/CLEOPR.2015.7375906]
19. Dubey, R.; Lahijani, B. V.; Kim, M.-S.; Barakat, E.; Häyrinen, M.; Roussey, M.; Kuittinen, M.; Herzig, H. P. Near-field investigation of Bloch surface wave based 2D optical components. Proceeding of SPIE Photonics West 10106, San Francisco, CA, USA, 2017. [10.1117/12.2250394]
20. Kim, M.-S.; Lahijani, B. V.; Descharmes, N.; Straubel, J.; Negredo, F.; Rockstuhl, C.; Häyrinen, M.; Roussey, M.; Kuittinen, M.; Herzig, H. P. Subwavelength focusing of Bloch surface waves. *ACS Photon.* **2017**. *4*, 1477–1483. [10.1021/acsphotonics.7b00245]
21. Doskolovich, L. L.; Bezus, E. A.; Bykov, D. A.; Soifer, V. A. Spatial differentiation of Bloch surface wave beams using an on-chip phase-shifted Bragg grating. *J. Opt.* **2016**, *18*, 115006. [10.1088/2040-8978/18/11/115006]
22. Dubey, R.; Barakat, E.; Herzig, H. P. Bloch Surface Waves Based Platform for Integrated Optics. Proceeding of IEEE Photonics Conference, 1092-8081, Reston, Virginia, USA, 2015. [10.1109/IPCon.2015.7323451]
23. Herzig, H. P.; Barakat, E.; Dubey, R.; Kim, M.-S. Optics in 2D Bloch surface wave phenomena and applications. 15th Workshop on Information Optics (WIO), 1-3, Barcelona, Spain, 2016. [10.1109/WIO.2016.7745573]
24. Dubey, R.; Barakat, E.; Kim, M.-S.; Herzig, H. P. Near-field characterization of 2D disk resonator on Bloch surface wave platform. 14th International Conference on Near-field Optics, Nanophotonics, and Related Techniques, Hamamatsu, Japan, 2016. [infoscience.epfl.ch/record/221501]
25. Herzig, H. P.; Barakat, E.; Yu, L.; Dubey, R. Bloch surface waves, a 2D platform for planar optical integration. 13th Workshop on Information Optics (WIO), Neuchatel, Switzerland, 2014. [10.1109/WIO.2014.6933280]
26. Dubey, R.; Barakat, E.; Herzig, H. P. Near Field Investigation of Bloch Surface Based Platform for 2D Integrated Optics. PIERS Progress In Electromagnetics Research Symposium, Prague, Czech Republic, 2015. [infoscience.epfl.ch/record/210317]
27. Kim, M.-S.; Dubey, R.; Barakat, E.; Herzig, H. P. Exotic optical elements generating 2D surface waves. EOS Topical Meeting on Trends in Resonant Nanophotonics, Berlin, Germany, 2016. [infoscience.epfl.ch/record/221568]
28. Dubey, R.; Barakat, E.; Herzig, H. P. Bloch Surface Based Platform for Optical Integration. TOM 5 – Metamaterials, Photonic Crystals and Plasmonics: Fundamentals and Applications, Berlin, Germany, 2014. [infoscience.epfl.ch/record/207822]
29. Khan, M. U.; Corbett, B. Bloch surface wave structures for high sensitivity detection and compact waveguiding. *Sci. Technol. Adv. Mater.* **2016**, *17*, 398-409. [10.1080/14686996.2016.1202082]
30. Kovalevich, T.; Boyer, P.; Suarez, M.; Salut, R.; Kim, M.S.; Herzig, H.P.; Bernal, M.P.; Grosjean, T. Polarization controlled directional propagation of Bloch surface wave. *Opt. Express*. **2017**, *25*, 5710-5715. [10.1364/OE.25.005710]
31. Kovalevich, T.; Ndao, A.; Suarez, M.; Tumenas, S.; Balevicius, Z.; Ramanavicius, A.; Baleviciute, I.; Häyrinen, M.; Roussey, M.; Kuittinen, M.; Grosjean, T.; Bernal, M.P. Tunable Bloch surface waves in anisotropic photonic crystals based on lithium niobate thin films. *Opt. Lett.* **2016**, *41*, 5616-5619. [10.1364/OL.41.005616]

32. Kovalevich, T.; Kim, M.S.; Belharet, D.; Robert, L.; Herzig, H.P.; Grosjean, T.; Bernal, M.P. Experimental evidence of Bloch surface waves on photonic crystals with thin film LiNbO3 as a top layer. *Photonics Res.* **2017**, *5*, 649-653. [10.1364/PRJ.5.000649]
33. Sreekanth, K.V.; Zeng, S.; Shang, J.; Yong, K.-T.; Yu, T. Excitation of surface electromagnetic waves in a graphene-based Bragg grating. *Sci. Rep.* **2012**, *2*, 737. [10.1038/srep00737]
34. Liang, G.; Neophytou, N.; Nikonov, D. E.; Lundstrom, M. S. Performance projections for ballistic graphene nanoribbon field-effect transistors. *IEEE Trans. Electron Devices*. **2007**, *54*, 677–682. [10.1109/TED.2007.891872]
35. Schuler, S.; Schall, D.; Neumaier, D.; Dobusch, L.; Bethge, O.; Schwarz, B.; Krall, M.; Mueller , T. Controlled Generation of a p−n Junction in a Waveguide Integrated Graphene Photodetector. *Nano Lett.* **2016**, *16*, 7107−7112. [10.1021/acs.nanolett.6b03374]
36. Zhou , Y.; Wang, C.; Xu D.-H.; Fan, R.-H.; Zhang, K.; Peng, R.-W.; Hu, Q.; Wang, M. Tuning the dispersion relation of a plasmonic waveguide via graphene contact. *EPL.* **2014**, *107*, 34007 (1-5). [10.1209/0295-5075/107/34007]
37. Marchena, M.; Janner, D.; Chen, T. L.; Finazzi, V.; Pruneri, V. Low temperature direct growth of graphene patterns on flexible glass substrates catalysed by a sacrificial ultrathin Ni film. *Opt. Mater. Express*. **2016**, *6*, 2487-2507. [10.1364/OME.6.002487]
38. Tien, P. K.; Ulrich, R.; Martin, R. J. Modes of propagating light waves in thin deposited semiconductor films. *Appl. Phys. Lett.* **1969**, *14*, 291–294. [10.1063/1.1652820]
39. Gassenq, A.; Hattasan, N.; Cerutti, L.; Rodriguez, J. B.; Tournié, E.; Roelkens, G. Study of evanescently-coupled and grating assisted GaInAsSb photodiodes integrated on a silicon photonic chip. *Opt. Express*. **2012**, *20*, 11665-11672. [10.1364/OE.20.011665]